\documentclass[twocolumn]{aastex631}

\usepackage{times,amsmath}
\usepackage[T1]{fontenc}

\hypersetup{pdfauthor={},
            pdftitle={},
            pdfkeywords={},
            bookmarksnumbered=true}
\pdfoutput=1

\usepackage[utf8]{inputenc}
\usepackage{graphicx}
\usepackage{amssymb}
\usepackage{amsmath}
\usepackage{float}
\usepackage{multirow}
\usepackage{textcomp}
\usepackage{gensymb}
\usepackage{enumitem}  
\usepackage{hyperref}
\usepackage{natbib}
\usepackage{comment}
\usepackage{systeme}
\usepackage[Symbol]{upgreek}
\usepackage{xcolor}
\usepackage{xspace}
\definecolor{xlinkcolor}{cmyk}{1,1,0,0}

\newcommand{\hi}{\textsc{H$\,$i}\xspace}

\shorttitle{SYMPHANY}
\shortauthors{Deb et al.}

\begin{document}

\title{SYMPHANY- SYnergy of Molecular PHase And Neutral hYdrogen in galaxies in A2626}

\correspondingauthor{Tirna Deb}
\email{tirna.deb@cfa.harvard.edu}

\author[0000-0003-1078-2539]{Tirna Deb}
\affiliation{Center for Astrophysics \textbar\ Harvard \& Smithsonian, 60 Garden St, Cambridge, MA 02138, USA}

\author{Garrett K. Keating}
\affiliation{Center for Astrophysics \textbar\ Harvard \& Smithsonian, 60 Garden St, Cambridge, MA 02138, USA}

\author{Nikki Zabel}
\affiliation{Department of Astronomy, University of Cape Town, Private Bag X3, Rondebosch 7701, South Africa}

\author{Alessia Moretti}
\affiliation{INAF-Padova Astronomical Observatory, Vicolo dell’Osservatorio 5, I-35122 Padova, Italy}

\author{Cecilia Bacchini}
\affiliation{DARK, Niels Bohr Institute, University of Copenhagen, Jagtvej 155, 2200 Copenhagen, Denmark}

\author{Timothy A Davis}
\affiliation{Cardiff Hub for Astrophysics Research \& Technology, School of Physics \& Astronomy, Cardiff University, Queens Buildings, Cardiff CF24 3AA, UK}

\author{Bianca M. Poggianti}
\affiliation{INAF-Padova Astronomical Observatory, Vicolo dell’Osservatorio 5, I-35122 Padova, Italy}

\author{Marco Gullieuszik}
\affiliation{INAF-Padova Astronomical Observatory, Vicolo dell’Osservatorio 5, I-35122 Padova, Italy}

\author{Benedetta Vulcani}
\affiliation{INAF-Padova Astronomical Observatory, Vicolo dell’Osservatorio 5, I-35122 Padova, Italy}

\author{Yara Jeffé}
\affiliation{Departamento de Física, Universidad Técnica Federico Santa María, Avenida España 1680, Valparaíso, Chile}

\affiliation{Millennium Nucleus for Galaxies (MINGAL)}

\author{Neven Tomičić}
\affiliation{Department of Physics, Faculty of Science, University of Zagreb, Bijenička 32, 10000 Zagreb, Croatia}

\author{Toby Brown}
\affiliation{Herzberg Astronomy and Astrophysics Research Centre, National Research Council of Canada, 5071 West Saanich Rd, Victoria, BC, V9E 2E7, Canada}

\begin{abstract}

We present an analysis of the molecular and atomic gas properties of 10 spatially resolved galaxies in the A2626 cluster (z = 0.055), observed as part of the SYMPHANY project. Using CO(2–1) observations from ALMA and SMA, together with \hi data from MeerKAT, we examine the interplay between gas phases and environmental influences. A joint morphological and kinematic analysis reveals that while the kinematic behavior of \hi and CO are often similar, morphologically the atomic gas is more extended compared to centrally concentrated CO distributions, making it more susceptible to environmental stripping.  However, we also find evidence for molecular gas asymmetries, disturbed velocity fields, and H$_2$ deficiencies in some galaxies, indicating that H$_2$ reservoirs can also be disrupted in dense environments. Compared to Virgo—a dynamically unrelaxed, non-cool-core cluster—A2626’s relaxed, cool-core structure likely results in less intense ram-pressure stripping. This may allow galaxies to retain more atomic gas, while local interactions or pre-processing may still affect the molecular phase, causing relatively low \hi deficiencies but high H$_2$ deficiencies in A2626 galaxies. This is also reflected in the higher gas fractions, slightly elevated SFRs, along with shorter depletion timescales (0.3–3 Gyr), implying moderately enhanced star formation efficiency A2626 galaxies. Moreover, the lack of correlation between H$_2$ deficiency and cluster-centric distance or velocity suggests that molecular gas evolution in A2626 may be shaped more by early infall or local interactions than by current cluster location.

\end{abstract}




\section{Introduction} \label{sec:intro}

The morphology-density relation \citep{Oemler1974, Dressler1980} for galaxies highlights the profound influence of cosmic environment on galaxy morphology, and star formation (SF) activity. Dense environments, such as galaxy groups and clusters, accelerate the quenching of SF and facilitate the transformation of gas-rich, late-type galaxies into gas-poor, early-type systems. This transformation is driven by both gravitational perturbations \citep{Tinsley1979, Moore1996} and hydrodynamical processes \citep{Gunn1972, Cowie1977, Balogh2000}, such as ram pressure stripping (RPS,  \citealt{Solanes2001, chung2009}). While gravitational interactions affect both stars and the interstellar medium (ISM), hydrodynamical processes primarily impact the ISM \citep{Gunn1972}. However, the effectiveness of these mechanisms in different cosmic environments, and their relative impacts on the ISM constituents and star formation rate (SFR), remain active areas of investigation.

 \renewcommand{\arraystretch}{1}
\begin{table}
    \centering
    \caption{Properties of A2626 \citep{deb2023}.}
    \begin{tabular}{lclc}\hline 
   Environment & cz ($km/s$) & $\sigma$ ($km/s)$ & R$_{200}$ (Mpc)                 \\ \hline 
 A2626  & 16576 &660 $\pm$ 26  & 1.59\\
     \hline
    \end{tabular}
    \label{tab:cluster_properties}
\end{table}

\begin{figure}
    \centering
        \includegraphics[width=0.5\textwidth]{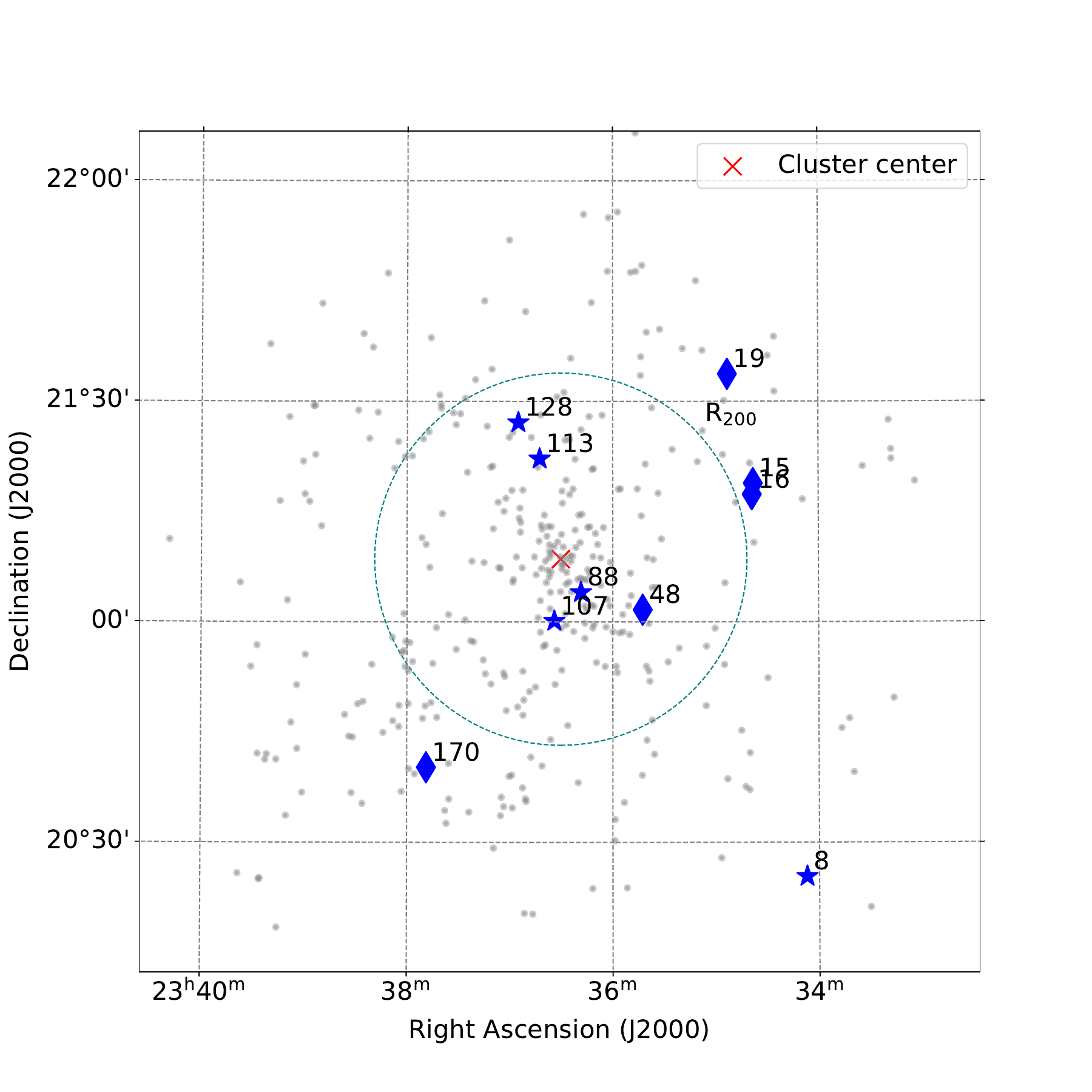}
 \caption{Sky distribution of galaxies in A2626. Star and diamond markers denote non-substructure and substructure galaxies respectively, targeted by ALMA and/or SMA within A2626. Small grey points represent other cluster galaxies with optical redshifts, while the dotted circle indicates the cluster's R$_{200}$ radius.}
     \label{fig:Sky_dist_a2626}
\end{figure} 

Observations of atomic hydrogen (\hi) in galaxies reveal that \hi reservoirs are significantly affected by cluster environments, often resulting in \hi deficiencies compared to field galaxies \citep{Solanes2001, chung2009, loni2021, Hess2022, Serra2023}. In contrast, the centrally concentrated and tightly gravitationally bound molecular gas (\(\mathrm{H}_2\)) is less susceptible to environmental influences (e.g. \citealt{Bacchini2023}). Nevertheless, studies of \(\mathrm{H}_2\) in cluster galaxies yield varying results across clusters \citep{Kenney1989, Brown2021, Zabel2019, Zabel2022, Zabel2024}. Given that \(\mathrm{H}_2\) is the primary fuel for star formation, understanding its response to dense environments is crucial for unraveling the broader picture of galaxy evolution.

\setlength{\tabcolsep}{10pt} 
\renewcommand{\arraystretch}{1.3} 
\begin{table*}[!htbp]
\caption{Observational overview: ALMA and SMA observations.}
\begin{center}
\begin{tabular}{lcc} 
\hline\hline
Observation Details      & ALMA (Band 6)         & SMA (230 GHz)       \\ 
\hline
Observing Date           & Oct, 2022            & Sep-Oct, 2021   \\ 
Configuration          & C2            & Subcompact   \\ 
Frequency Range        & 217.3--219.2 GHz         & 211--223 GHz      \\ 
LAS (Largest Angular Scale)        & 10$\arcsec$         & 25$\arcsec$     \\
Channel Width (MHz) & 0.488               & 17.875            \\ 
Velocity Width (km/s) & 0.6 & 23 \\
Synthesized beam size & 1.5$\arcsec \times 1.1\arcsec$ & 4.6$\arcsec \times 3.9\arcsec$ \\ 
Primary beam FWHM & 26$\arcsec$ & 52$\arcsec$ \\
RMS (noise)       & 2.8-6.6 mJy/beam         & 20-28 mJy/beam      \\

\hline
\end{tabular}
\end{center}
\label{Table:obs}
\end{table*}

Current insights into the environmental impact on the cold gas-phase ISM are predominantly derived from studies of low-redshift clusters such as Virgo \citep{Kenney1989, chung2009, Brown2021}, Fornax \citep{Zabel2019, loni2021, Zabel2024}, Coma \citep{Boselli1997, Casoli1996, Molnar2021}, and the WIde-field Nearby Galaxy-cluster Survey (WINGS, z=0.04-0.07) cluster sample observed by the GAs Stripping Phenomena in galaxies (GASP) survey \citep{Poggianti2017, jaffe2018, PoggiantiNature2017}. The Virgo Environment Traced in CO (VERTICO) survey mapped molecular gas in 51 Virgo Cluster galaxies using the Atacama Large Millimeter/submillimeter Array (ALMA) to study how dense environments disrupt gas disks, shaping star formation and galaxy evolution \citep{Brown2021}. 
The GASP survey studied the different gas phases (atomic, molecular, ionised), star formation properties of 100 stripping candidate galaxies (z = 0.04–0.07) across various environments, aiming to investigate gas removal processes, their timescales, efficiency, and impact on star formation \citep{Poggianti2017, jaffe2018}. In Virgo, galaxies exhibit significant \hi deficiencies \citep{chung2009} but only moderate \(\mathrm{H}_2\) deficiencies \citep{Brown2021, Zabel2022}. Interestingly, \(\mathrm{H}_2\) suppression is observed in filamentary structures feeding into Virgo \citep{Castignani2022}. Fornax galaxies are deficient in both \hi and \(\mathrm{H}_2\) \citep{Zabel2019, loni2021, Serra2023}. In GASP galaxies with star-forming tails and Virgo jellyfish galaxies, the total  \(\mathrm{H}_2\) content is higher, or these galaxies exhibit lower  \(\mathrm{H}_2\) deficiency, compared to undisturbed field galaxies of similar stellar mass or other galaxies in Virgo \citep{Zabel2022, Moretti2023}. This suggests that ram pressure stripping plays a key role in removing gas from galaxies, leading to the depletion of their star-forming fuel and ultimately driving their quenching, or in some cases, partially converting \hi into \(\mathrm{H}_2\). These findings underscore the complex interplay of environmental processes, as \hi-deficient galaxies somehow retain their star-forming \(\mathrm{H}_2\) gas \citep{Moretti2020a, Moretti2023}. What is the dominant environmental mechanism affecting star-forming cold gas  (\hi and \(\mathrm{H}_2\))? How does the cluster environment influence the cold gas content of infalling galaxies? How do these processes influence the baryon cycle in galaxies? Do the properties of the galaxy cluster change the answers to the proceeding questions?

Expanding our understanding of the environmental impact on the multi-phase ISM across various regions of more clusters—from the core to the infalling outskirts—is essential for addressing these questions.

\subsection{The A2626 Galaxy Cluster: A Case Study} \label{sec:sample}

The galaxy cluster A2626, located at a redshift of \( z=0.055 \), is a moderately massive system (\(\sim 5 \times 10^{14} \, M_\odot\)) with a velocity dispersion of \( \sigma = 660 \pm 26 \, \mathrm{km/s} \) and a virial radius (R$_{200}$) of 1.59 Mpc \citep{HealySS2021, deb2023}. A large cosmic volume centered on A2626 ($\sim$ 2 R$_{200}$) was surveyed using MeerKAT, revealing three distinct overdensities representing diverse cosmic environments \citep{HD2021}. The galaxies in the A2626 cluster span the redshift range \( 0.0475 \leq z \leq 0.0615 \).

MeerKAT \hi observations \citep{HD2021, deb2023} reveal a clear \hi deficiency towards the cluster center and in substructures. Several ``jellyfish" galaxies, showing signs of ongoing RPS, further highlight that environmental effects are active and important in A2626 \citep{deb2022}. These observations demonstrate asymmetric, offset, and truncated \hi discs in galaxies located both in the cluster core and infalling substructures. Substructure identification, based on the Dressler-Shectman (DS test, \citealt{Dressler1988}) applied to optical spectroscopy, reveals a diverse sample for studying environmental impacts \citep{HealySS2021}.

Photometric imaging from the DECam Legacy Survey (DECaLS; \citealt{dey2019}) and Sloan Digital Sky Survey (SDSS; \citealt{York2000, Aguado2018}) complement these \hi observations (\hi mass limit $\sim 2 \times 10^8$ M$_\odot$, \citealt{HD2021}). Stellar masses and star formation rates (SFRs) were derived using WISE photometry (T. Jarrett, private communication). The stellar masses were computed from the 3.4$\mu$m (W1) and 4.6$\mu$m (W2) emission, while the SFRs were estimated based on the PAH emission in the 12$\mu$m (W3) band. For further details, see \citet{cluver2014, cluver2017} and \citet{jarrett2013, jarett2019}.

The sample includes 76 galaxies detected in the MeerKAT pointing centered on A2626. Two primary galaxy populations are identified:
\begin{enumerate}
    \item \textbf{Non-substructure galaxies:} A population of 57 galaxies in the high-density environment of A2626 lacking identifiable substructures.
    \item \textbf{Substructure galaxies:} A population of 34 galaxies residing in substructures, such as groups within the cluster environment, identified using the DS test on optical spectroscopic cluster members \citep{HealySS2021}.
\end{enumerate}

This volume-limited survey, with its diverse \hi- and SFR-selected sample, provides a unique opportunity to investigate the environmental impacts on the multi-phase ISM across different cosmic environments within the A2626 cluster.

\setlength{\tabcolsep}{10pt} 
\renewcommand{\arraystretch}{1.2} 
\begin{table*}[t]
\caption{List of all relevant HI and stellar physical and morphological characteristics of the observed galaxies.}
\begin{center}
\begin{tabular}{ccccccccccccccc}
\hline
\multicolumn{1}{c}{HI ID} &
\multicolumn{1}{c}{Name} &
\multicolumn{1}{c}{SS} &
\multicolumn{1}{c}{z} &
\multicolumn{1}{c}{Log(M$\rm_{HI}$)} &
\multicolumn{1}{c}{$\pm$} &
\multicolumn{1}{c}{Log(M$_{*}$)} &
\multicolumn{1}{c}{$\pm$} &
\multicolumn{1}{c}{SFR} &
\multicolumn{1}{c}{$\pm$} &
\multicolumn{1}{c}{\hi Class} \\
& & 
\multicolumn{2}{c}{} &
\multicolumn{2}{c}{(M$_{\odot}$)} &
\multicolumn{2}{c}{(M$_{\odot}$)} &
\multicolumn{2}{c}{(M$_{\odot}$/yr)} &
\multicolumn{2}{c}{} &
\multicolumn{2}{c}{} &
\multicolumn{1}{c}{}\\
\multicolumn{1}{c}{(1)}&
\multicolumn{1}{c}{(2)}&
\multicolumn{1}{c}{(3)}&
\multicolumn{1}{c}{(4)}&
\multicolumn{1}{c}{(5)}&
\multicolumn{1}{c}{(6)}&
\multicolumn{1}{c}{(7)}&
\multicolumn{1}{c}{(8)}&
\multicolumn{1}{c}{(9)}&
\multicolumn{1}{c}{(10)}&
\multicolumn{1}{c}{(11)}\\
\hline
 8 & J233407.08+202522.2 & 0 & 0.0515 & 8.89 & 0.64 & 10.06 & 0.15 & 1.17 & 0.41 & 0\\
  15 & J233438.15+211851.7 & 1 & 0.0538 & 9.93 & 0.03 & 10.36 & 0.15 & 3.33 & 1.15 & 1\\
  16 & J233438.80+211721.0 & 1 & 0.0529 & 9.07 & 0.08 & 10.22 & 0.11 & 2.00 & 0.70 & 2\\
  19 & J233453.14+213344.9 & 1 & 0.0554 & 9.38 & 0.06 & 9.78 & 0.13 & 2.03 & 0.71 & 1\\
  48 & J233542.52+210140.8 & 1 & 0.0529 & 8.83 & 0.09 & 9.91 & 0.22 & 0.83 & 0.31 & 1\\
  88 & J233618.57+210402.6 & 0 & 0.0533 & 9.77 & 0.02 & 9.89 & 0.14 & 2.31 & 0.81 & 0\\
  107 & J233634.13+210007.4 & 0 & 0.0526 & 8.76 & 0.09 & 10.18 & 0.19 & 1.65 & 0.58 & 0\\
  113 & J233642.70+212212.0 & 0 & 0.0550 & 9.58 & 0.03 & 10.08 & 0.13 & 2.52 & 0.88 & 0\\
  128 & J233655.10+212708.9 & 0 & 0.0528 & 9.46 & 0.05 & 10.99 & 0.21 & 4.43 & 1.53 & 0\\
  170 & J233748.75+204013.3 & 1 & 0.0561 & 9.64 & 0.04 & 10.12 & 0.12 & 2.27 & 0.79 & 1\\
\hline
\end{tabular}
\end{center}
{Notes: Column (1): The assigned \hi identifier from \citealt{HD2021}. Column (2): The SDSS identifier for the optical counterpart of the \hi detected galaxies, based on their Right Ascension and Declination (J2000.0). Column (3): Substructure identifier. SS=`0',`1': non-substructure and substructure galaxies in A2626, SS=`2': galaxies in the Swarm. Column (4): Redshift measured as the midpoint of the 20\% line width of the \hi global profile. Column (5,6): Log of \hi mass and uncertainty as mentioned in \citealt{HD2021}. Column (7,8): Log of stellar mass and uncertainty were derived from WISE photometry, it was kindly provided by T. Jarrett (private communication). Column (9,10): 12 $\mu$m star formation rates star formation rate and uncertainty from WISE observations (provided by T. Jarrett, private communication). Column (11): \hi class based on \citep{Yoon2017}.}
\label{tab:targets}
\end{table*}

\section{CO Observations and Data Processing} \label{sec:data}

The SYnergy of Molecular PHase And Neutral hYdrogen in galaxies in A2626 (SYMPHANY) project aims to investigate the molecular gas phase within a sample of galaxies in A2626. The selected galaxies were observed in the CO(2-1) line using the Sub Millimeter Array (SMA) and/or the Atacama Large Millimeter/submillimeter Array (ALMA).  

All galaxies in the SYMPHANY sample had previously been imaged in \hi\ by MeerKAT, covering a range of \hi\ content \citep{deb2023}. Additionally, they were detected in radio continuum (also by MeerKAT) and selected based on their estimated CO disc size. The sample selection ensured that the CO disc would span at least three beams, assuming a conservative CO radius estimate of half the galaxy’s optical radius. Notably, some targets are located in the outskirts of the cluster (R $> R_{200}$), offering insights into galaxies in different stages of environmental interaction, including potential `pre-processed' systems (see Figure \ref{fig:Sky_dist_a2626}).

Among the 76 \hi-detected galaxies in A2626, a subset of 44 met these criteria. Of these, 17 were observed in ALMA Cycle 9, and 9 were detected, spanning a range of stellar masses (\(10^{9.8} - 10^{11} \, \mathrm{M_\odot}\)), star formation rates (\(0.83 - 4.43 \, \mathrm{M_\odot \, yr^{-1}}\)), \hi\ content (\(10^{8.8} - 10^{9.9} \, \mathrm{M_\odot}\)), and global as well as local environments (both substructure and non-substructure galaxies).  Additionally, from the list of 44 galaxies, 9 were proposed for SMA observations, of which 6 were observed between September–October 2021. Among these, 4 were detected by SMA, 3 of which were also detected by ALMA. Table \ref{tab:targets} provides the list of targets observed with ALMA and/or SMA, along with their optical coordinates, spectroscopic redshifts, stellar masses, star formation rates, and \hi classifications. The \hi morphology was categorized following the scheme of \citet{Yoon2017}, based on visual inspection of the stellar extents (from DECaLS color images), \hi disc extents, and the morphology of the outermost \hi contours as observed by MeerKAT \citep{HD2021}.

For this study, we utilized data reduced by the standard ALMA Cycle 9 pipeline, which performs automated calibration, imaging, and quality control to ensure science-ready data \citep{Hunter2023}. The pipeline achieves a flux calibration accuracy of 3–5\%, with uncertainties up to 10\% in suboptimal conditions \citep{Fomalont2014, alma_technical_handbook}.

Analysis of SMA data was reduced using the SMA COMPASS pipeline (Keating et al., in prep.), which performs spectral-based flagging (via looking for outliers in amplitude and chi-squared statistics over variable integration times) and baseline-based flagging (discarding baselines where little-to-no coherence is seen on calibrator targets), gains/bandpass calibration, and flux scaling using the Butler-JPL-Horizons 2012 \citep{2012ALMAM.594....1B} model of Uranus. Images were produced from the data and deconvolved performed using the CLEAN algorithm \citep{1974A&AS...15..417H}.  

For the ALMA data cubes, a 4$\sigma$ detection mask with a 90\% reliability threshold was generated using SoFiA \citep{Serra2015, Westmeier2021}, with the smooth+clip algorithm applied to create accurate masks for the CO(2-1) emission. SoFiA was run on the cube after binning 10 channels, where each original channel had a width of 0.64 km/s, resulting in a final channel width of 6.4 km/s and an improved signal-to-noise ratio by a factor of $\sqrt{10} \sim 3.2$. Binning preserves integrated flux, improves detectability, and maintains accurate velocity centroid and line width measurements, as the binned resolution remains well below the typical molecular gas velocity dispersion of $10 - 20$ km/s in cluster galaxies \citep{Bolatto2013, Dobbs2014}.
Gaussian smoothing kernels with Full Width at Half Maximum (FWHM) values of 0$\arcsec$, 0.63$\arcsec$, 1.26$\arcsec$, and 1.89$\arcsec$ were applied.

For the SMA data cubes, due to the high noise levels, with the criteria requiring a detection threshold of $3\sigma$ and a reliability of 90\%, SoFiA did not produce any reliable masks. Rather than loosening the criteria in SoFiA parameter settings and risking unreliable detections, a tailored masking approach was implemented with a dedicated script to accurately identify and isolate CO emission. This method involved generating low and high signal-to-noise masks based on $3\sigma$ and $5\sigma$ thresholds, respectively, with $\sigma$ representing the local noise. Connected regions were identified across spatial and spectral dimensions, and spurious detections were removed using a minimum pixel count criterion based on beam size. To recover faint emission near bright regions, binary dilation was applied to refine the final mask.


\begin{figure*}
\centering
\gridline{
  \fig{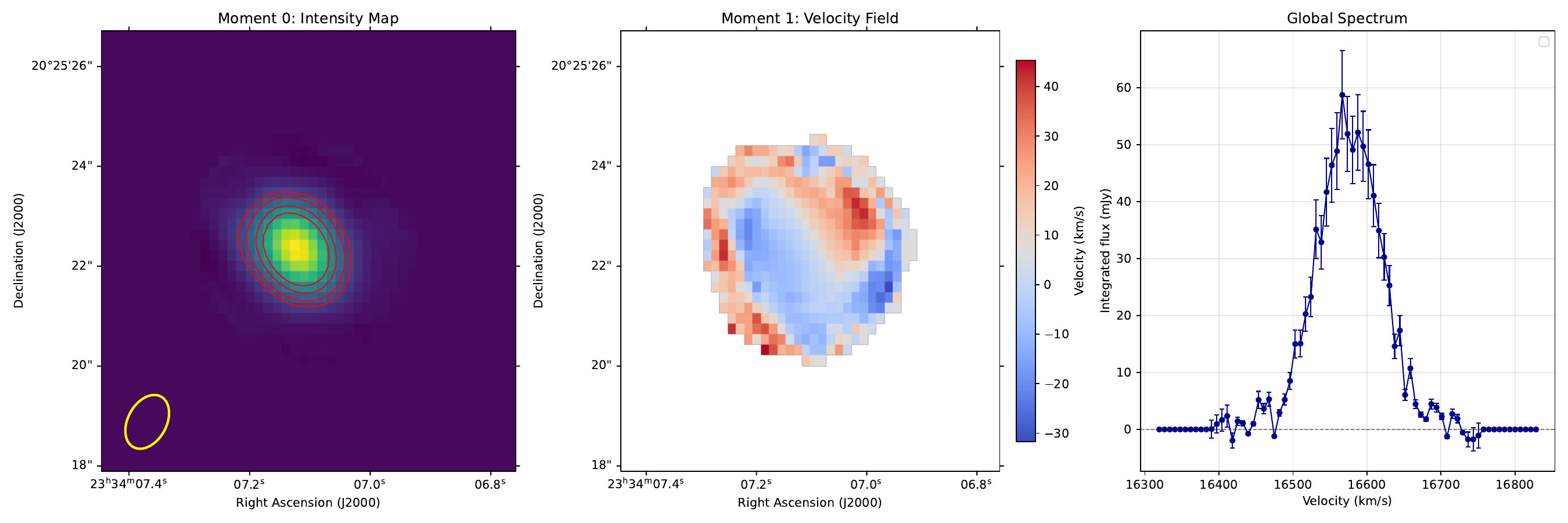}{0.95\textwidth}{(a) mom zero, velocity field and global profile of \hi ID 8.}
}
\gridline{
  \fig{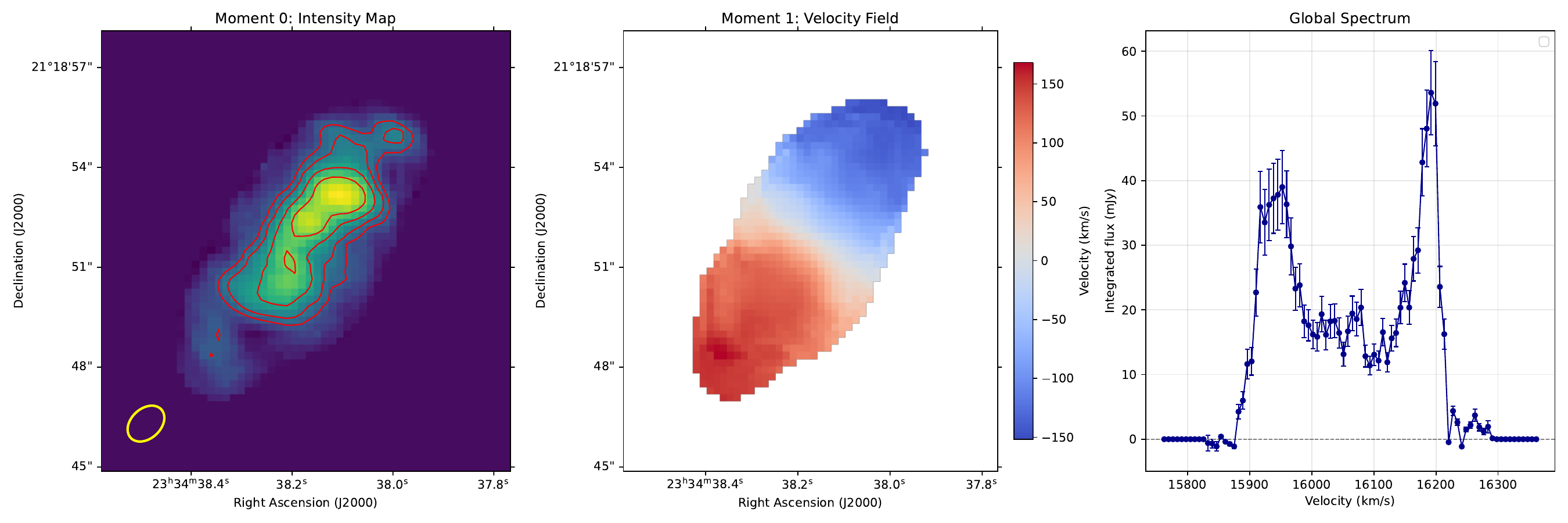}{0.95\textwidth}{(b) mom zero, velocity field and global profile of \hi ID 15.}
}
\gridline{
  \fig{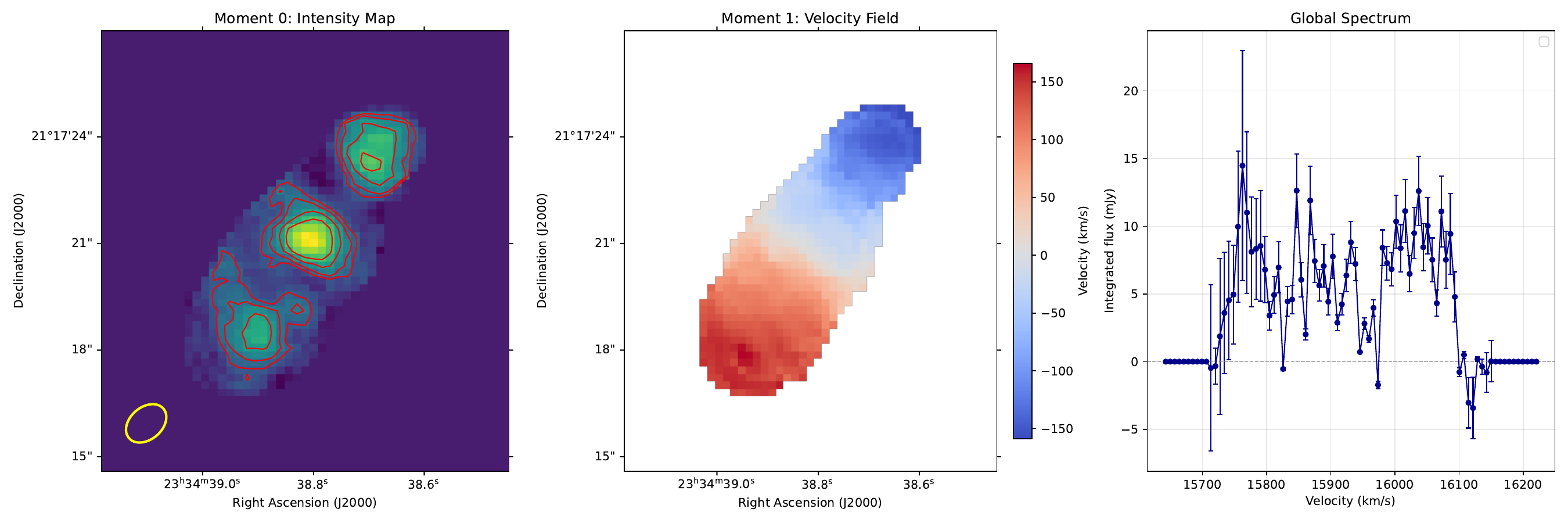}{0.95\textwidth}{(c) mom zero, velocity field and global profile of \hi ID 16.}
}
\caption{Overview of \hi kinematic and global profile properties (1/3). Each panel shows the moment zero map, velocity field, and global \hi profile for a different galaxy.}
\label{fig:combined_atlas_1}
\end{figure*}

\begin{figure*}
\centering
\gridline{
  \fig{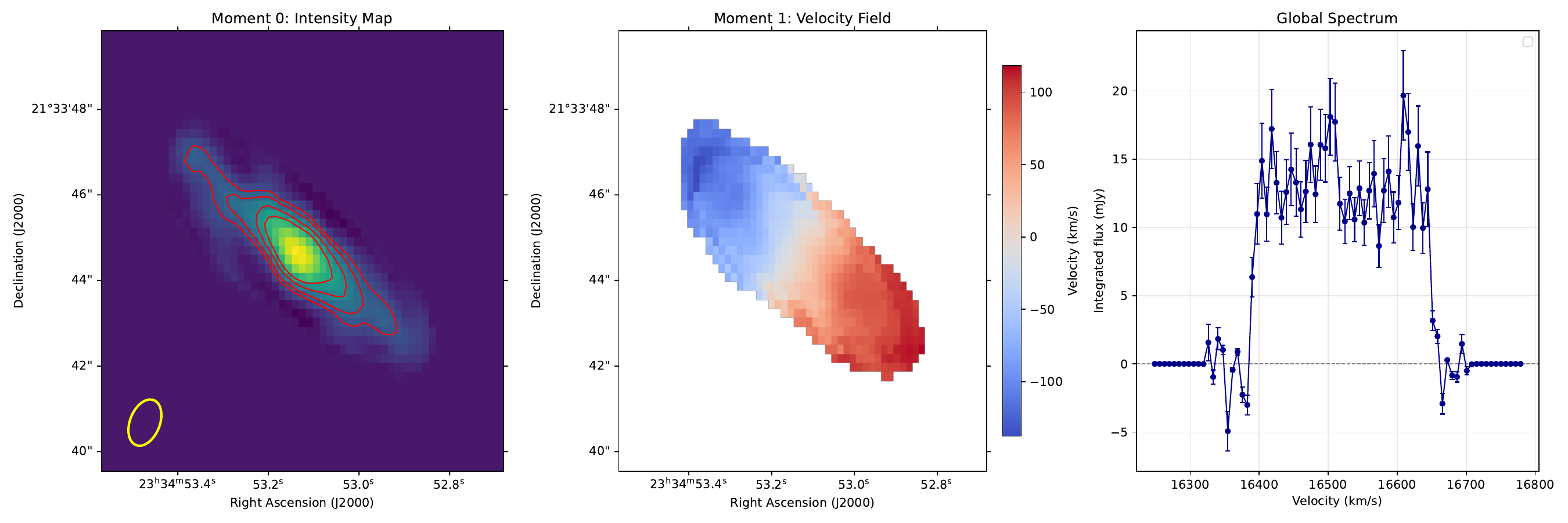}{0.95\textwidth}{(d) mom zero, velocity field and global profile of \hi ID 19.}
}
\gridline{
  \fig{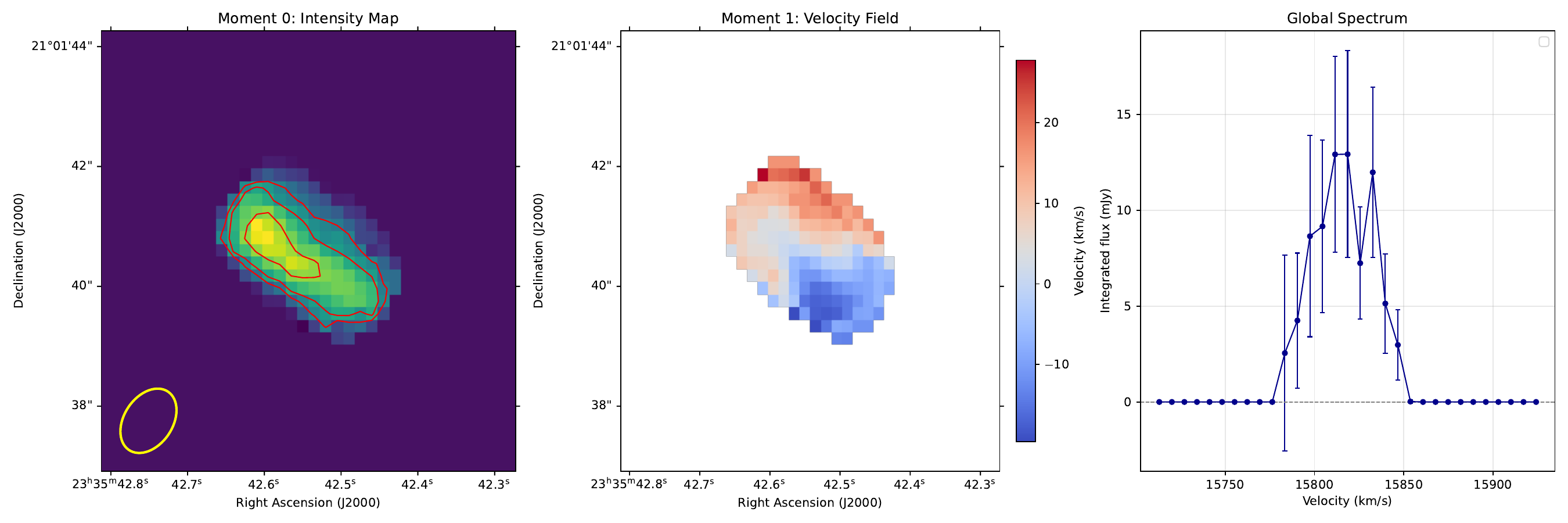}{0.95\textwidth}{(e) mom zero, velocity field and global profile of \hi ID 48.}
}
\gridline{
  \fig{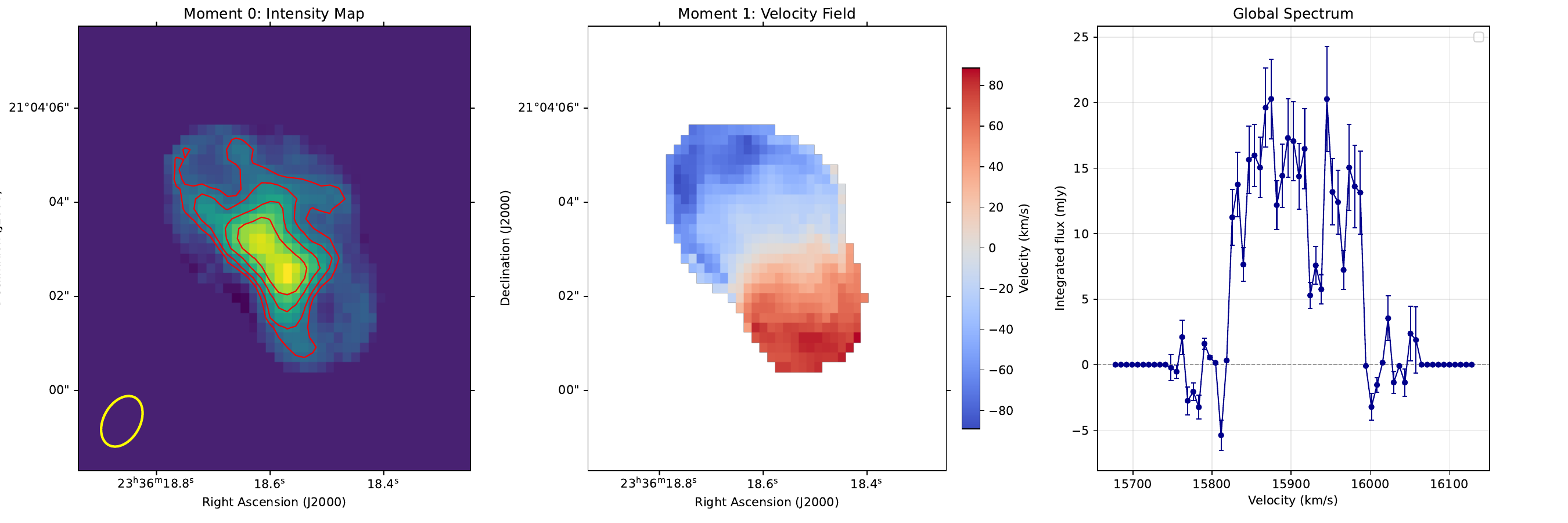}{0.95\textwidth}{(f) mom zero, velocity field and global profile of \hi ID 88.}
}
\caption{Overview of \hi kinematic and global profile properties (2/3).}
\label{fig:combined_atlas_2}
\end{figure*}

\begin{figure*}
\centering
\gridline{
  \fig{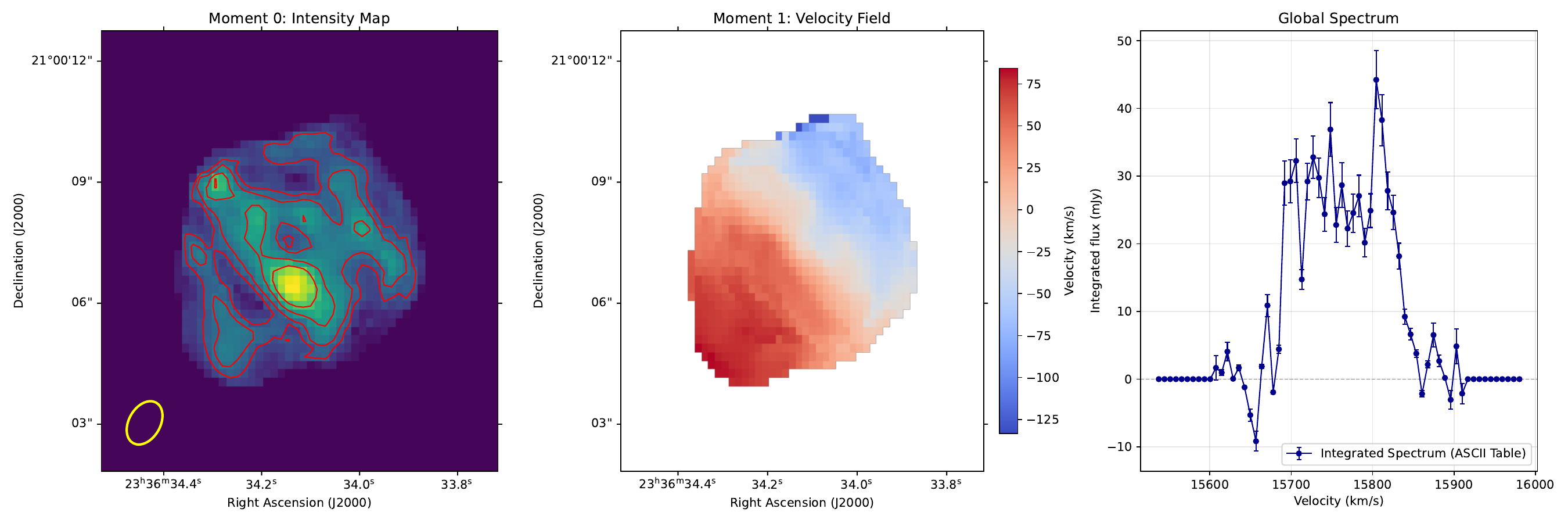}{0.95\textwidth}{(g) mom zero, velocity field and global profile of \hi ID 107.}
}
\gridline{
  \fig{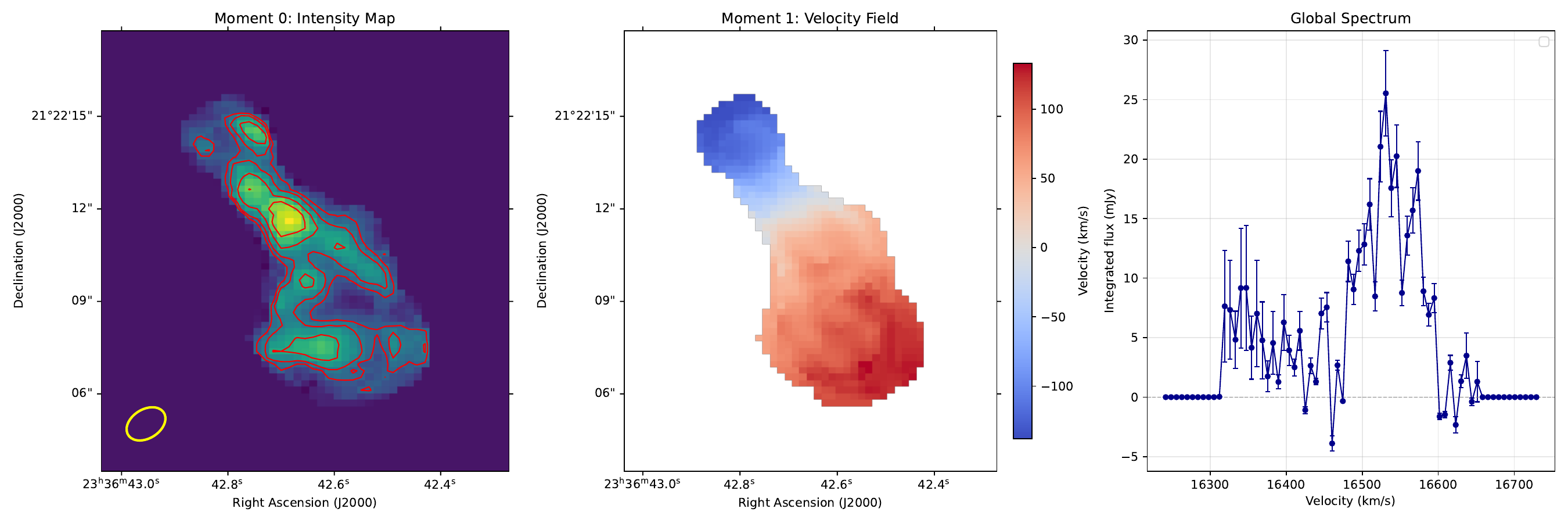}{0.95\textwidth}{(h) mom zero, velocity field and global profile of \hi ID 113.}
}
\gridline{
  \fig{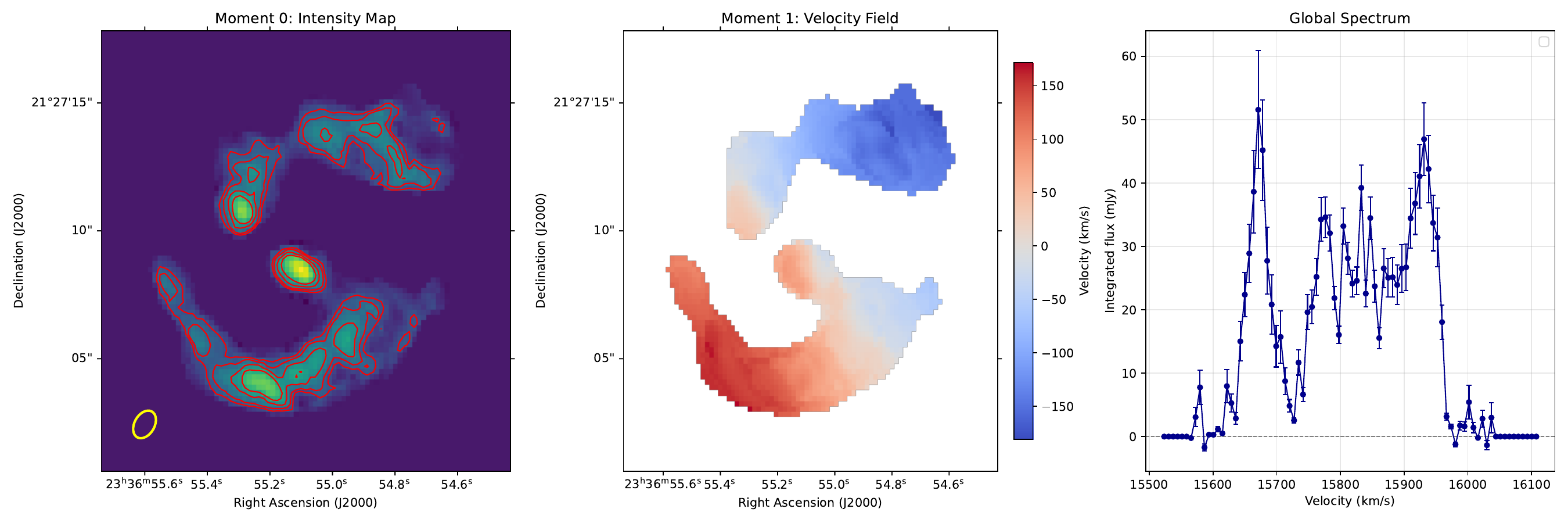}{0.95\textwidth}{(i) mom zero, velocity field and global profile of \hi ID 128.}
}
\caption{Overview of \hi kinematic and global profile properties (3/3).}
\label{fig:combined_atlas_3}
\end{figure*}

\setlength{\tabcolsep}{6pt} 
\renewcommand{\arraystretch}{1.2} 
\begin{table*}[t]
\caption{Comparison of CO fluxes and molecular gas masses derived from SMA and ALMA observations for selected galaxies in the A2626 cluster.}
\centering
\begin{tabular}{cccccccc}
\hline
\hline
\multicolumn{1}{c}{HI ID} &
\multicolumn{1}{c}{z} &
\multicolumn{1}{c}{freq (GHz)} &
\multicolumn{1}{c}{$f_{\rm sum, SMA}$} &
\multicolumn{1}{c}{$f_{\rm sum, ALMA}$} &
\multicolumn{1}{c}{log$_{10}$(M$_{\rm H_2, SMA}$)} &
\multicolumn{1}{c}{log$_{10}$(M$_{\rm H_2, ALMA}$)} \\
& & & \multicolumn{1}{c}{(Jy km/s)} & \multicolumn{1}{c}{(Jy km/s)} &
\multicolumn{1}{c}{(M$_\odot$)}  &
\multicolumn{1}{c}{(M$_\odot$)} \\
\hline
15 & 0.0538 & 218.77 & 8.50 & 9.42 & 9.21 &  9.25  \\
88 & 0.0533 & 218.87 & 8.01 & 5.70 & 9.18 &  9.04  \\
107 & 0.0526 & 219.02 & 9.26 & 8.51 & 9.25 &  9.21  \\
\hline
\end{tabular}
\label{tab:co_fluxes}
\end{table*}

\setlength{\tabcolsep}{5pt} 
\renewcommand{\arraystretch}{1.2} 
\begin{table*}[t]
\caption{Properties of ALMA and/or SMA-detected SYMPHANY galaxies in A2626. Here the measurements for all but galaxy 170 (SMA observation) are from ALMA observations. Columns (1): \hi ID, Columns (2): RMS noise level, Columns (3) \& (4): CO and \hi center offsets (in kpc), Columns (5) \& (6): \hi and H$2$ deficiencies, Columns (7) \& (8) : Log of molecular gas mass (M$_{\mathrm{H_2}}$) with associated uncertainties. Offsets indicate the displacement of the gas centers relative to the optical center.}
\centering
\begin{tabular}{cccccccc}
\hline
\multicolumn{1}{c}{HI ID} &
\multicolumn{1}{c}{RMS} &
\multicolumn{1}{c}{CO offset} &
\multicolumn{1}{c}{\hi offset} &
\multicolumn{1}{c}{\hi def} &
\multicolumn{1}{c}{H$_{2}$ def} &
\multicolumn{1}{c}{M$\mathrm{_{H2}}$} &
\multicolumn{1}{c}{$\pm$} \\
\multicolumn{1}{c}{} &
\multicolumn{1}{c}{mJy/beam} &
\multicolumn{2}{c}{(kpc)} &
\multicolumn{1}{c}{} &
\multicolumn{1}{c}{} &
\multicolumn{2}{c}{(M$_{\odot}$)} \\
\multicolumn{1}{c}{(1)}&
\multicolumn{1}{c}{(2)}&
\multicolumn{1}{c}{(3)}&
\multicolumn{1}{c}{(4)}&
\multicolumn{1}{c}{(5)}&
\multicolumn{1}{c}{(6)}&
\multicolumn{1}{c}{(7)}&
\multicolumn{1}{c}{(8)} \\
\hline
 8 & 6.16 & 0.617 & 14.047 & -0.56 & 0.55 &  8.96 & 0.048\\
  15 & 4.82 & 0.173 & 2.055 & 0.50 & 0.42 &  9.14 & 0.052\\
  16  & 3.29 & 0.296 & 5.846 & 0.37 & -0.01 &  8.58 & 0.122\\
  19  & 3.48 & 0.798 & 0.846 & 0.19 & 0.29 &  8.81 & 0.08\\
  48  & 4.81 & 1.137 & 4.267 & -0.22 & -0.42 &  7.93 & 0.544\\
  88  & 3.59 & 0.915 & 3.399 & 0.53 & 0.83 &  8.57 & 0.12\\
  107  & 4.18 & 1.252 & 3.28 & -0.41 & 0.19 &  8.84 & 0.066\\
  113  & 2.83 & 2.483 & 2.174 & -0.32 & 0.07 &  8.61 & 0.186\\
  128  & 6.63 & 0.612 & 2.752 & 0.43 & -0.24 &  7.67 & 0.986\\
  170 & 25 & 2.215 & 2.815 & 0.01 & 0.29 &  8.74 & 0.352 \\
\hline
\end{tabular}
\label{tab:co_props}
\end{table*}


\begin{figure*}[t!]
    \centering
 {
    \includegraphics[width=\textwidth]{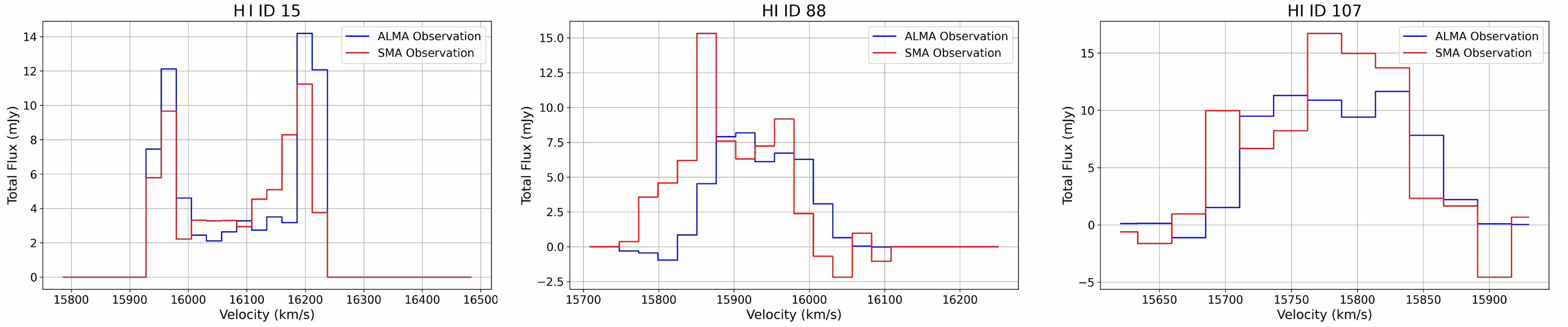} 
  } 
  \caption{Comparison of ALMA and SMA global profiles for galaxies in A2626. Both profiles have been matched to the same spatial and spectral resolution by smoothing the ALMA cube to the SMA resolution. After this adjustment, the profiles appear similar, demonstrating consistency in CO emission detections between the two facilities at comparable resolutions.} 
  \label{fig:A2626_ALMA_SMA_2025_GP} 
 \end{figure*}

\section{Results and Discussion}

\subsection{Description of individual galaxies}

Figure \ref{fig:combined_atlas_1} presents the moment 0 and moment 1 maps, along with the CO(2–1) global spectra, for all ALMA-detected galaxies in A2626. Each panel displays the molecular gas distribution or moment 0 (left panel), the velocity field  or moment 1 (middle panel), and the corresponding CO emission line profile (right panel), allowing a comparative view of variations in line width, flux peak, and kinematic structure. Moment 0 contours are drawn at levels of 0.7, 1, 1.5, and 2 times $\sigma_{\mathrm{map}}$, where $\sigma_{\mathrm{map}}$ is the root-mean-square of non-zero pixel intensities and serves as a characteristic scale for the integrated emission. The ALMA synthesized beam size is shown in the bottom-left corner of each image. Moment 1 maps use a velocity scale that shows redshifted and blueshifted regions across the rotating CO discs. The CO global profiles were generated using flux density measurements and corresponding errors from the SoFiA pipeline, summing the emission within SoFiA-generated masks for each velocity channel. These profiles were uniformly extracted, with error bars indicating measurement uncertainties. Differences in profile shapes reflect variations in molecular gas distribution and kinematics among the galaxies.

\hi ID 8: The galaxy is located significantly outside the cluster's virial radius ($\sim$2.2 R$_{200}$). In Figure \ref{fig:combined_atlas_1}, \hi ID 8 exhibits a rotating molecular gas disc traced via CO(2-1) emission. The morphology (moment zero) and kinematics (velocity field) indicate that the galaxy is relatively dynamically stable, with no strong signs of major perturbations. The CO intensity distribution and the presence of an overall velocity gradient suggest that the molecular gas is likely rotating and may be fueling ongoing star formation. The symmetric, single-peaked global spectrum supports this interpretation.

\hi ID 15: \hi ID 15 is a substructure galaxy, located just outside the virial radius, shows clear evidence of a rotating molecular gas disc, as indicated by the smooth velocity gradient in the moment 1 map. However, the moment 0 map reveals an elongated and asymmetric CO(2–1) intensity distribution, with the peak intensity offset from the geometric center. This asymmetry is echoed in the global CO spectrum, which exhibits a double-horned profile—characteristic of rotating discs—but with unequal peak heights and widths, suggesting kinematic differences between the two sides. The moment 0 contours trace regions of enhanced molecular gas surface brightness, emphasizing the lopsided nature of the disc.

\hi ID 16: \hi ID 16 is a substructure galaxy, also located just outside the virial radius, exhibits a rotating molecular gas disc, as indicated by the smooth velocity gradient in the moment 1 map, with a clear transition from redshifted to blueshifted velocities along the major axis. The moment 0 map shows a clumpy and extended CO(2–1) intensity distribution, with multiple peaks aligned along the disc, suggesting a non-uniform gas structure possibly shaped by internal dynamical processes or past interactions. The global CO spectrum is irregular and noisy, lacking a clearly defined double-horned profile, which may reflect the disturbed gas morphology or low signal-to-noise in parts of the disc. These features collectively point to a kinematically rotating but morphologically complex system.

\hi ID 19: \hi ID 19 is a substructure galaxy, located outside cluster radius, exhibits a rotating molecular gas disc, as indicated by the smooth and symmetric velocity gradient in the moment 1 map. The moment 0 map shows a compact elliptical CO(2–1) distribution aligned with the kinematic major axis, consistent with disc-like morphology. However, the global CO spectrum lacks a clearly defined double-horned or flat-topped shape, likely due to a combination of intrinsic gas asymmetries and limited signal-to-noise. While the velocity field suggests ordered rotation, the spectral profile makes it challenging to fully assess the dynamical state of the system.

\begin{figure*}[t!]
    \centering
 {
    \includegraphics[width=\textwidth]{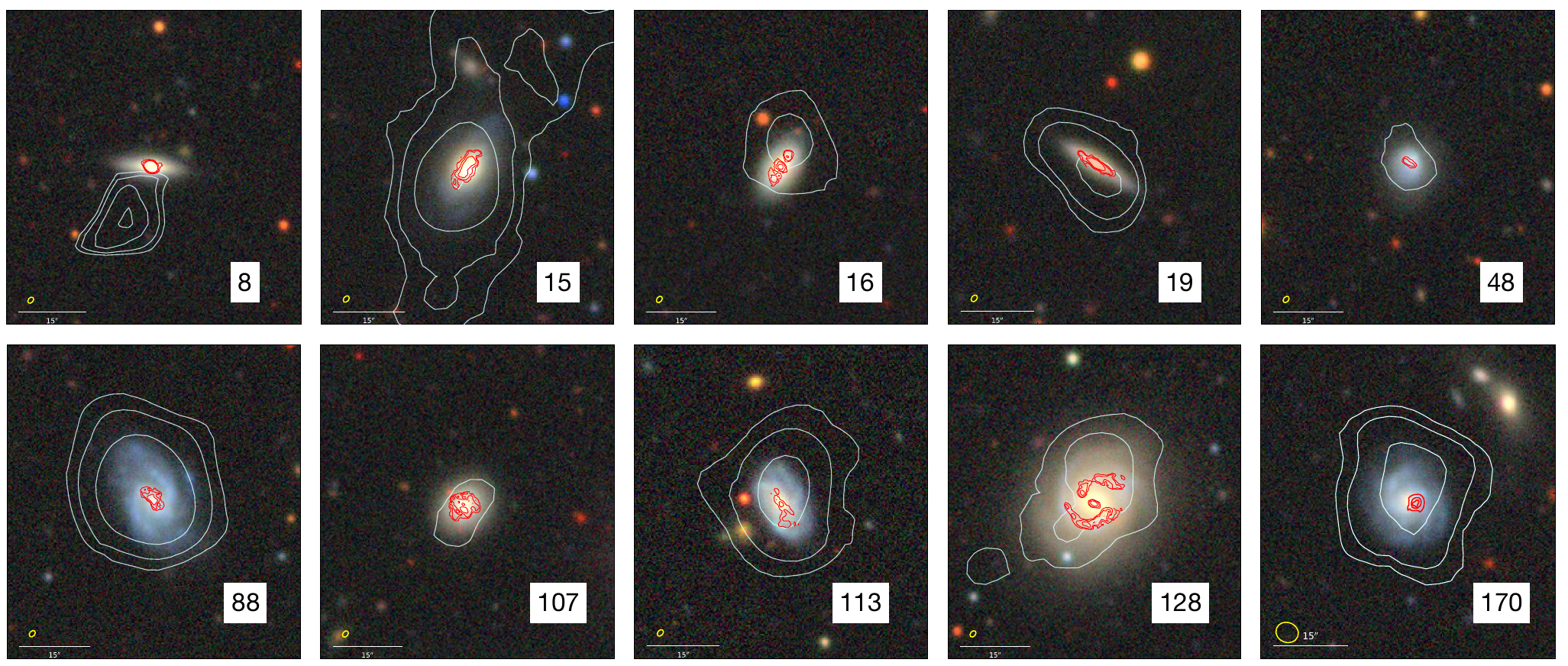} 
  } 
  \caption{SYMPHANY galaxies in the A2626 galaxy cluster, with CO (red) and \hi (white) contours overlaid on DECaLS RGB color images. CO and \hi observations are from ALMA and MeerKAT observations respectively (except \hi ID 170 is observed by SMA). A 15$\arcsec$ scale bar in the bottom-left corner of each image indicates the size of the \hi beam. The ALMA beam size is also shown. \hi contours are plotted at levels of $17 \times (2^n)$ where $n = [0, 1, 2,..]$, in units of cm$^{-2}$. CO contours are drawn at $[1, 2, 3, 4, ...] \times 5.0\sigma_{\rm rms}$. The lowest contour levels are $0.02$, $0.03$, $0.02$, $0.02$, $0.02$, $0.019$, $0.02$, $0.03$, $0.024$, and $0.025$ Jy beam$^{-1}$ for galaxies 8, 15, 16, 19, 48, 88, 107, 113, 128, and 170 respectively. CO discs appear centrally concentrated compared to the extended \hi distributions of the galaxies. } 
  \label{fig:SYMPHANY_overlays} 
 \end{figure*}

\hi ID 48: \hi ID 48 is a substructure galaxy, located within the cluster radius, and exhibits a compact molecular gas disc. The moment 0 map reveals an elliptical CO(2–1) intensity distribution with a bright region offset from the center and a CO tail extending towards the south-east. The moment 1 map shows a modest but coherent velocity gradient, which aligns with the direction of the CO extension and is broadly consistent with rotation. The global CO spectrum displays a single-peaked, asymmetric profile with relatively low signal-to-noise, lacking the double-horned signature—possibly due to a nearly face-on orientation, as inferred from the optical image (Figure \ref{fig:SYMPHANY_overlays}).

\hi ID 88: \hi ID 88 is a non-substructure galaxy, located close to the cluster center ($\sim$ 0.2 R$_{200}$), and exhibits a centrally concentrated but asymmetric molecular gas disc, with the moment 0 map showing a compact yet lopsided CO(2–1) intensity distribution. The moment 1 map reveals a velocity gradient across the disc that appears broadly consistent with rotation, though with significant asymmetries. The global CO spectrum displays a broad, moderately asymmetric double-peaked profile, characteristic of a rotating system.

\hi ID 107: \hi ID 107 is a non-substructure galaxy, located close to the cluster core ($\sim$ 0.6 R$_{200}$), exhibits a clumpy and extended molecular gas distribution, as seen in the moment 0 map, with multiple emission peaks spread across the disc. The moment 1 map shows a clear and continuous velocity gradient from redshifted to blueshifted regions, consistent with large-scale rotation. The global CO spectrum displays a broad, double-peaked profile with noticeable asymmetry and substructure, suggesting a rotating disc with non-uniform gas distribution or local kinematic disturbances. 

\hi ID 113: \hi ID 113 is a non-substructure galaxy, located within to the cluster radius ($\sim$ 0.2 R$_{200}$), presents an extended and asymmetric molecular gas distribution, as shown in the moment 0 map, with multiple clumps and a clear elongation along the kinematic axis. The moment 1 map reveals a strong and continuous velocity gradient, indicating large-scale rotation across the disc. The global CO spectrum displays a broad, asymmetric double-peaked profile with fluctuations in flux across the velocity range, consistent with an inclined, rotating disc experiencing either internal instabilities or environmental effects. The combination of kinematic regularity and morphological complexity suggests a dynamically evolving system.

\hi ID 128: \hi ID 128 is a spiral galaxy with clumpy molecular gas concentrated in the inner spiral arms, as seen in the moment 0 map. The CO(2–1) emission is distributed in distinct knots tracing the spiral structure near the center, while the moment 1 map reveals a coherent velocity gradient consistent with disc rotation. The global CO spectrum displays a broad, multi-peaked profile, shaped by the clumpy gas distribution and varying line-of-sight velocities across the arms. 

To estimate the molecular gas mass (M$_{\rm H_2}$, see Table \ref{tab:co_props}), we used the CO line flux following the equation from \citet{Watson2017}:

\begin{equation}
\left(\frac{M_{\rm H_2}}{M_{\odot}}\right) = 3.8 \times 10^3 
\left(\frac{\alpha_{10}}{4.3}\right) 
\left(\frac{r_{21}}{0.7}\right)^{-1} 
\left(\int S_{21}dv \right) (D_L)^2,
\end{equation}

where $\alpha_{10}$ is the CO-to-H$2$ conversion factor in M${_\odot}$ pc$^{-2}$ (K km s$^{-1}$)$^{-1}$, $r_{21}$ is the CO(2–1)/CO(1–0) flux ratio, $\int S_{21}dv$ is the integrated CO(2–1) line flux in units of Jy km s$^{-1}$, and $D_L$ is the luminosity distance in Mpc. We adopted $\alpha_{10} = 4.3$ from \citet{Bolatto2013}, the standard Milky Way value, corresponding to a conversion factor of $2 \times 10^{20}$ cm$^{-2}$ (K km s$^{-1}$)$^{-1}$, including the helium correction. For consistency with previous studies \citep{Moretti2018, Moretti2020a, Moretti2020b}, we used $r_{21} = 0.79$.

\subsection{ALMA vs SMA CO global profiles}\label{sec:co_profile}

The global velocity profiles or integrated spectra of \hi and CO gas provide critical insights into the gas distribution and kinematics in A2626 galaxies.

First, we compared the spectra of three galaxies (\hi ID 15, 88, and 107) for which we have both ALMA and SMA observations. To enable a direct channel-by-channel comparison of the CO global profiles, the ALMA cube was first smoothed to match the beam size and velocity channel width of the SMA cube. This ensured that both datasets had comparable spatial and spectral resolutions. A source mask was generated by running SoFiA on the smoothed ALMA cube, leveraging the higher signal-to-noise ratio of ALMA detections due to its superior sensitivity. The ALMA data cubes were then regridded to align with the spatial and spectral grid of the SMA cube, ensuring consistent velocity sampling. The ALMA-derived mask was applied to the SMA cube while maintaining alignment using the source center determined by SoFiA. This approach allowed us to extract global profiles from both cubes using the same number of pixels in the detection mask, ensuring a  consistent comparison of total flux and profile shapes across velocity channels. Table \ref{tab:co_fluxes} presents the comparison of ALMA and SMA CO fluxes and molecular gas masses for galaxies 15, 88, and 107 in the A2626 cluster.

Comparison of ALMA and SMA global profiles and H$_{2}$ masses shows that CO emission measurements are consistent within the uncertainties. This comparison demonstrates that ALMA does not resolve out significant flux compared to SMA, as the detected sources have sizes of within 10$\arcsec$, which is comparable to ALMA’s largest angular scale (LAS) of 10$\arcsec$ and well within SMA’s LAS of 25$\arcsec$. Given the respective rms values of 2.8–6.6 mJy/beam for ALMA and 20–28 mJy/beam for SMA, this confirms that ALMA is not missing extended CO emission due to spatial filtering.

\begin{figure*}[t!]
    \centering
 {
    \includegraphics[width=\textwidth]{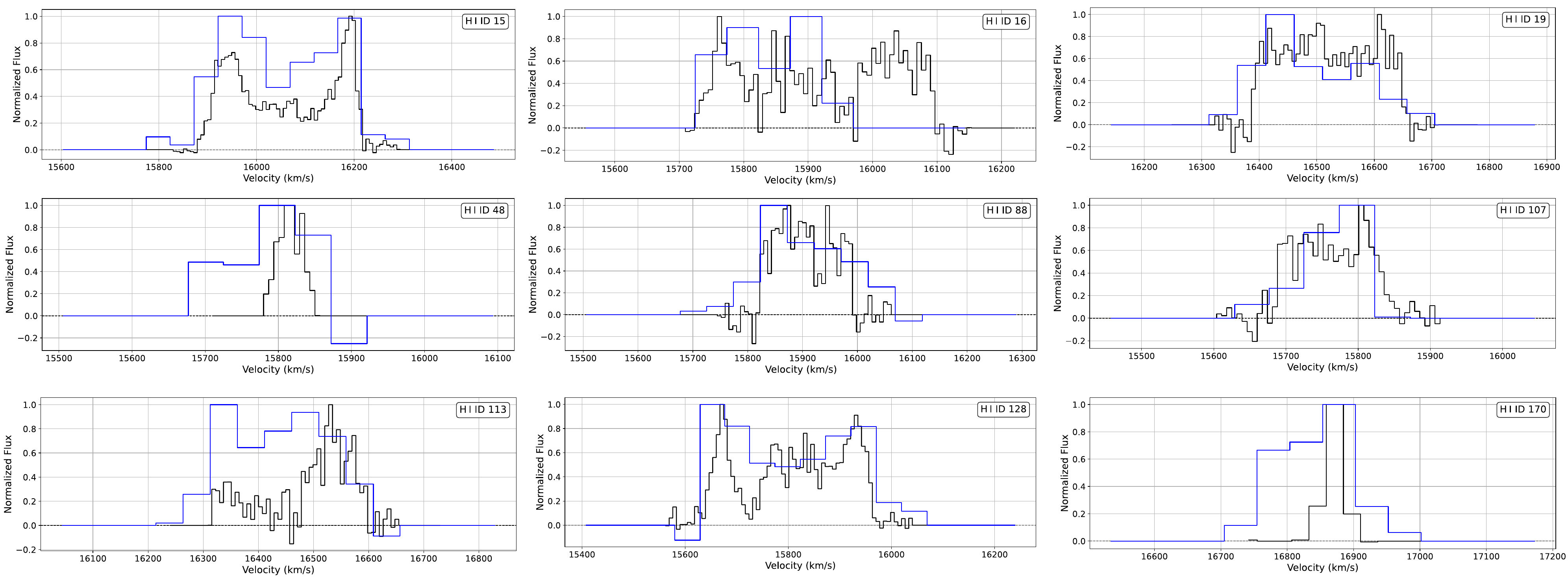} 
  } 
  \caption{Global CO (black) and \hi (blue) profiles for A2626 galaxies. All but one (SMA observations of \hi ID 170) CO profiles are based on ALMA observations. The x-axis represents the velocity in km/s, while the y-axis normalized flux density for galaxies. the \hi and CO components generally span similar ranges, suggesting that both gas phases trace the overall galactic rotation. In some cases (e.g., \hi ID 48 and 170), the \hi velocity width exceeds that of CO,  possibly due to the presence of extended, disturbed atomic gas in the outer regions, while the molecular gas remains more centrally concentrated and less affected.} 
  \label{fig:A2626_HI_CO_global_profiles} 
 \end{figure*}

\subsection{Multi-phase Interplay of Atomic and Molecular Gas}\label{sec:multiphase_interplay}

To investigate the environmental effects on the cold interstellar medium (ISM) in cluster galaxies, we analyze spatially resolved overlays and global profiles of atomic (\hi) and molecular (CO) gas for a sample of 10 A2626 galaxies observed with ALMA, SMA and MeerKAT. The galaxies exhibit a diversity of morphological and kinematic features in both gas phases, reflecting not only intrinsic differences in galaxy structure, but also varying degrees of environmental processing.

Figure~\ref{fig:SYMPHANY_overlays} presents the spatial overlays of \hi (white contours) and CO (red contours) on DECaLS RGB images. The \hi contours, derived from MeerKAT observations \citep{HD2021, deb2023}, are generally more extended and asymmetric compared to the centrally concentrated CO emission from ALMA. Contours are plotted at standardized significance levels: \hi at $17 \times (2^n)$ cm$^{-2}$ and CO at $[1, 2, 3, ...] \times 5.0\sigma_{\rm rms}$. A 15\arcsec\ scale bar indicates the \hi circular beam size, and the ALMA beam is shown for reference.

 Figure~\ref{fig:A2626_HI_CO_global_profiles} shows the corresponding global profiles of \hi (blue) and CO (black) flux as a function of velocity. These profiles provide comparative insights into the gas kinematics and extent, complementing the spatial information. For most galaxies, the CO emission remains centrally concentrated, while the \hi distribution is often more extended and asymmetric—showing tails or offsets—indicative of environmental processes such as ram-pressure stripping (RPS) or tidal interactions. In terms of velocity distribution, the \hi and CO components generally span similar ranges, suggesting that both gas phases trace the overall galactic rotation. However, in a few cases (e.g., \hi ID 48 and 170), the \hi velocity width exceeds that of CO, possibly due to disturbed atomic gas in the outskirts that is not matched by the more centrally confined molecular component. Below, we describe individual systems that exhibit noteworthy features in both the overlays and global profiles.

\hi ID 8: The \hi detection lies outside the MeerKAT primary beam, suggesting a potential misclassification in \hi, that is why it is not in included in Figure \ref{fig:A2626_HI_CO_global_profiles}. The CO and optical centers are well aligned. 

\hi ID 15: The \hi distribution is much extended compared to the centrally located CO distribution. \hi morphology shows an offset and an asymmetry in the spatial distribution as well as in the global profile. The CO emission is slightly asymmetric in the same direction (see both Figure \ref{fig:combined_atlas_1} and \ref{fig:SYMPHANY_overlays}), suggesting plausible tidal interactions with the neighbouring galaxy.

\hi ID 16: The \hi is one-sided and significantly offset compared to the CO, likely due to RPS/ tidal interactions. The global \hi profile lacks emission on the redshifted side where the CO peak lies, indicating efficient \hi stripping.

\hi ID 19: The \hi shows a slight spatial offset from the CO and optical centers. Its broader, asymmetric global profile may be linked to tidal interaction within the substructure population or RPS.

\hi ID 48: Located in the cluster core, both \hi and CO show slight offsets in the same direction, possibly indicating the early effects of gas stripping or interaction with the substructure population. The \hi profile is more extended than the CO profile, suggesting a disrupted atomic gas reservoir, while the molecular component remains relatively concentrated and less affected.

\hi ID 88: \hi appears asymmetric and slightly offset from the optical center, while CO is also mildly displaced. The global profiles support this, showing a broader, slightly asymmetric \hi distribution.

\hi ID 107: The \hi detection is weak and asymmetric, both spatially and spectrally. CO is more symmetric but slightly offset. The combined evidence suggests advanced gas stripping, with CO beginning to respond.

\hi ID 113: Both \hi and CO are offset in the same direction and aligned with the stellar morphology. This galaxy exhibits the most asymmetric CO profile, with significant differences between the redshifted and blueshifted parts, suggesting external disturbances.

\hi ID 128: A highly \hi-deficient galaxy. CO is asymmetric and extended because of the spiral structure, with its emission concentrated in the upper part, mirroring the \hi offset (see both Figure \ref{fig:combined_atlas_3} and \ref{fig:SYMPHANY_overlays}).

\hi ID 170: Located near the cluster boundary, the \hi is broader and offset, while the CO appears narrow and slightly displaced. The low signal-to-noise SMA detection may have captured only part of the CO reservoir.

Overall, the data illustrate that \hi is typically more extended and environmentally sensitive than CO, based on the previous observations (e.g. \citealt{Brown2021, Moretti2019}). While \hi asymmetries and displacements are common, several galaxies also show signs of CO disruption, indicating progressive gas removal affecting both phases. In fact, many galaxies in the A2626 sample exhibit markedly asymmetric or truncated CO discs—particularly among the larger spirals—suggesting that environmental processes are significantly depleting their molecular reservoirs as well.


 \begin{figure*}[t!]
    \centering
 {
    \includegraphics[width=0.8\textwidth]{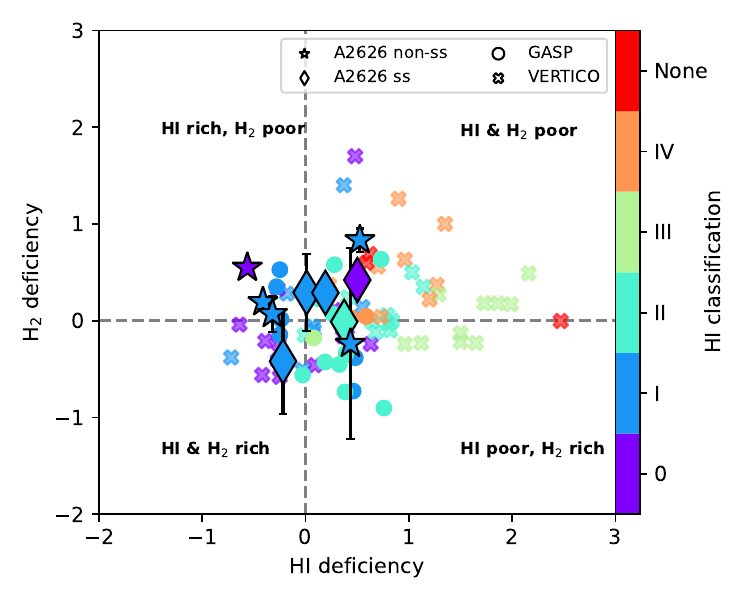} 
  } 
  \caption{Correlation between \hi and H${_2}$ deficiencies of SYMPHANY galaxies compared to Virgo \citep{Zabel2022} and three other cluster samples \citep{Moretti2023}. The ALMA-observed galaxies are color-coded based on classifications derived from optical and \hi morphologies \citep{Yoon2017}. Notably, no class III or IV galaxies are detected in both \hi and CO for the A2626 sample. The A2626 galaxies are comparatively less \hi and H${_2}$ deficient, which reflect environmental or selection effects.} 
  \label{fig:ALMA_galaxies_defHI_defH2} 
 \end{figure*}

\subsection{\hi vs H$_{2}$ Deficiency in A2626 Galaxies}\label{sec:hi_h2_def}

The balance between atomic (\hi) and molecular (H$_2$) gas in cluster galaxies provides critical insight into the impact of environmental processes on their interstellar medium (ISM). Previous studies have shown that massive galaxies undergoing peak stripping often exhibit enhanced H$_2$/\hi ratios compared to field galaxies due to preferential removal of \hi while retaining molecular gas \citep{Moretti2022}. In the Virgo cluster, \citet{Zabel2022} found a weak correlation between \hi and H$_2$ deficiencies, suggesting that \hi loss does not necessarily imply molecular gas depletion. Virgo jellyfish galaxies, in particular, appear less H$_2$ deficient than other cluster members. Similarly, \citet{Moretti2020a, Moretti2023} showed that GASP galaxies with star-forming tails are slightly \hi poor but relatively H$_2$ rich compared to field galaxies. These results imply that ram-pressure stripping (RPS) affects \hi and H$_2$ on different timescales, with molecular gas being more gravitationally bound and resilient to immediate removal \citep{Lee2017, Boselli2022, Bacchini2023}.

To compare the cold gas properties of SYMPHANY galaxies with Virgo and GASP systems, we classified A2626 galaxies based on visual asymmetries and gas content using criteria from \citet{Yoon2017}, based on \hi morphology, extent, and deficiency (classification details are provided in the last column of Table 1). Their system includes galaxies with one-sided \hi features but untruncated disks (Class I), highly asymmetric and truncated \hi disks (Class II), and symmetric but severely truncated \hi disks with high deficiencies (Class III). Additional classes capture more mildly truncated disks with lower surface densities (Class IV) and galaxies that do not fit into the previous categories, such as field-like spirals or those showing signs of tidal interaction (Class V). We quantified \hi and H$_2$ deficiencies using the stellar mass-based definition from \citet{Zabel2022}:

\begin{equation}
    \mathrm{def}_i = \log M_{i,\mathrm{exp}} - \log M_{i,\mathrm{obs}},
\end{equation}

where $i = $ \hi or H$_2$, $M_{i,\mathrm{exp}}$ is the expected mass at a given stellar mass (based on xGASS/xCOLDGASS control samples), and $M_{i,\mathrm{obs}}$ is the observed mass. We used 10 logarithmically spaced stellar mass bins to calculate the median expected masses and ensure consistent comparison.

Figure~\ref{fig:ALMA_galaxies_defHI_defH2} presents the \hi and H$_2$ deficiencies of SYMPHANY galaxies alongside Virgo and GASP galaxies. SYMPHANY galaxies are color-coded by their \hi class, determined from optical and \hi morphologies.

Notably, our sample lacks class III and IV galaxies, which typically show strongly disturbed, truncated \hi discs. Although selected to span a range of \hi properties, SYMPHANY galaxies generally exhibit lower \hi and H$_2$ deficiencies compared to Virgo and GASP samples, clustering near the origin (no deficiency). However, for a given \hi deficiency, the SYMPHANY galaxies tend to show elevated H$_2$ deficiencies, suggesting that they are relatively \hi-normal but comparatively H$_2$-poor. This pattern could indicate environmentally driven molecular gas depletion, or possibly reflect sensitivity limitations in detecting lower surface brightness CO emission in some galaxies. Overall, this suggests that environmental effects in A2626 are more subtle and possibly selective, with fewer galaxies showing strong atomic gas depletion, but some evidence for molecular gas suppression.

\begin{figure}[t!]
    \centering
 {
    \includegraphics[width=0.5\textwidth]{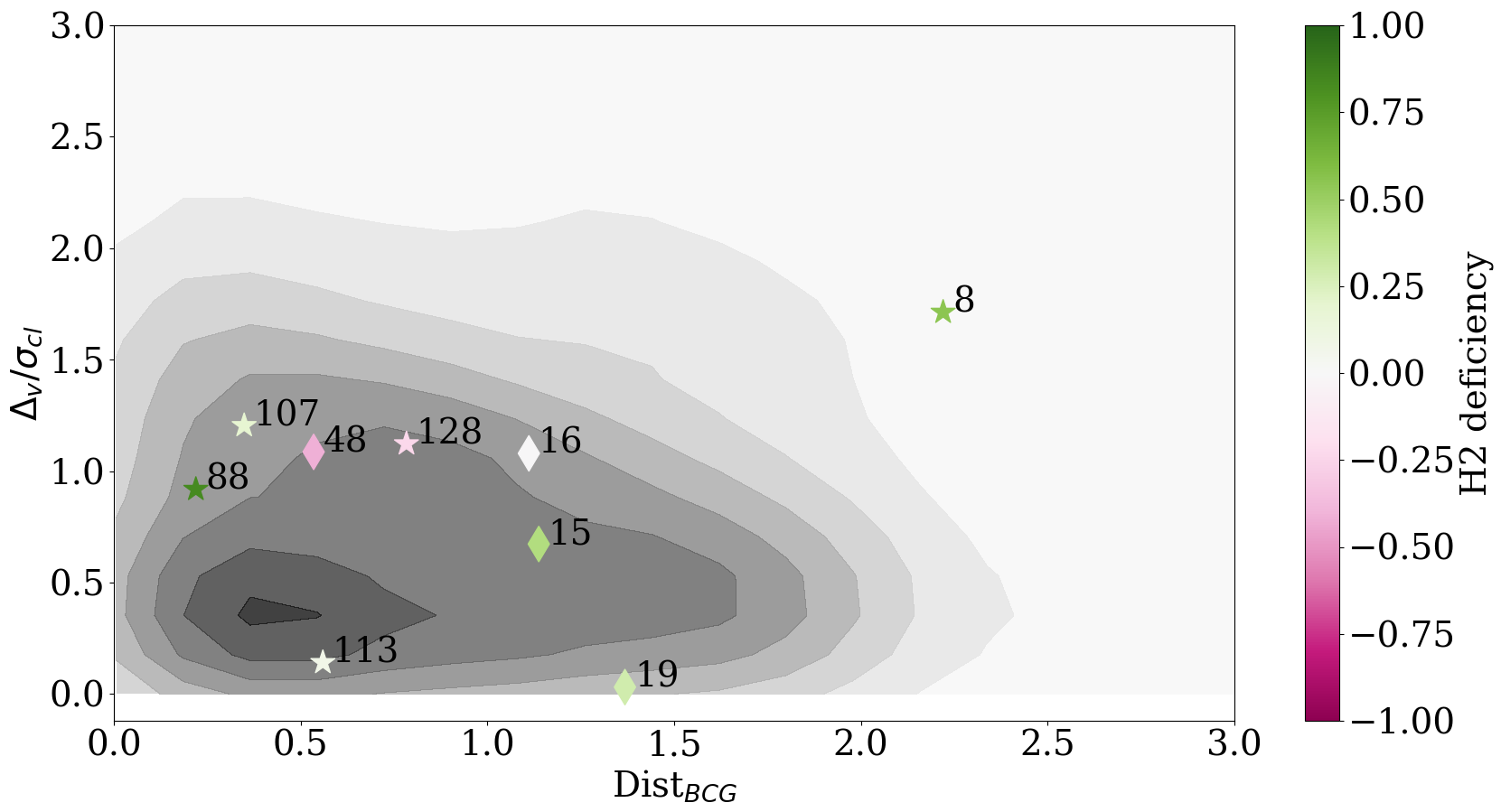} 
  } 
  \caption{The distribution of SYMPHANY galaxies in the projected phase-space of A2626. The x-axis shows the projected clustercentric distance normalized by R$_{200}$, and the y-axis shows the line-of-sight velocity relative to the cluster mean, normalized by the cluster velocity dispersion. Grayscale contours represent the galaxy number density distribution. Diamond and star markers denote non-substructure and substructure galaxies, respectively, with marker colors representing the H$_{2}$ deficiencies of the galaxies. No clear correlation is observed between the environment, morphology, and H$_{2}$ deficiency. } 
  \label{fig:ALMA_H2_def_PS} 
 \end{figure}

To further probe the dynamical state and gas processing in A2626, we constructed a projected phase-space diagram (Figure~\ref{fig:ALMA_H2_def_PS}). The x-axis shows the clustercentric distance normalized by R$_{200}$, and the y-axis represents the galaxy’s line-of-sight velocity offset from the cluster mean, normalized by the cluster velocity dispersion ($\Delta v / \sigma_{\mathrm{cl}}$, \citealt{HD2021, deb2023}). Background greyscale contours represent the galaxy density distribution in A2626 based on redshifts from \citet{HealySS2021}, and colored markers denote the H$_2$ deficiency of each galaxy. Stars represent non-substructure galaxies, and diamonds represent substructure members.

No strong correlation is observed between H$_2$ deficiency and projected clustercentric distance or velocity. This may indicate that H$_2$ content increases during early infall stages—before being consumed or stripped (as also seen in \citealt{Moretti2023}) —and that local interactions or preprocessing may play a more critical role than cluster-centric location alone in shaping molecular gas properties in A2626.

\begin{figure}
    \centering
 {
    \includegraphics[width=0.5\textwidth]{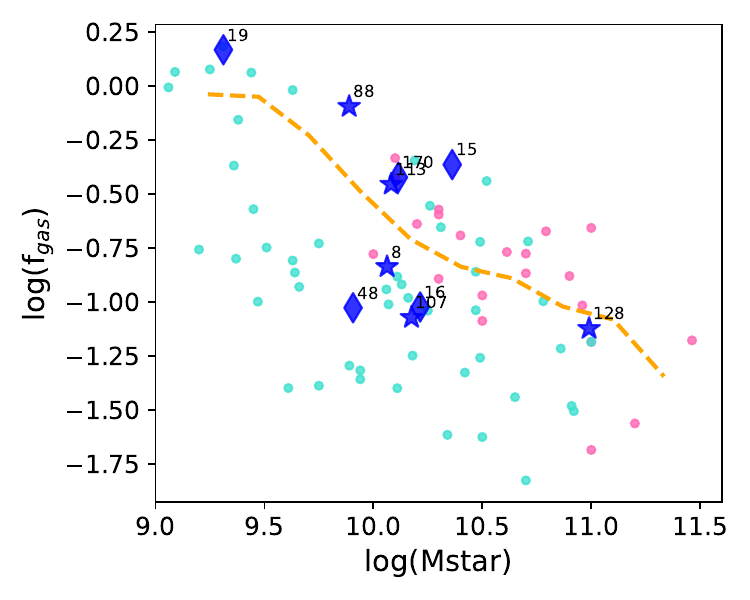} 
  } 
  \caption{Total gas (\hi+H$_{2}$) fraction as a function of stellar mass for A2626 galaxies (large blue markers), Virgo galaxies (turquoise markers), and GASP galaxies (pink markers). In A2626, star markers represent non-substructure galaxies, while diamond markers denote substructure galaxies. The orange dotted line indicates the median \hi+H$_{2}$ mass for binned stellar masses of xCOLDGASS field galaxies.} 
  \label{fig:ALMA_galaxies_fgas_Mstar} 
 \end{figure}

\subsection{Total Gas Fraction as a Function of Stellar Mass}\label{sec:gas_frac_stellar_mass}

To further investigate the impact of the cluster environment on the gas content of SYMPHANY galaxies, we examine how their total gas fraction ($\frac{M_{\rm HI} + M_{\mathrm H_2}}{M_{*}}$) compares with galaxies in other clusters and field environments. This comparison provides insight into how environmental processes influence the overall gas reservoirs of galaxies in dense regions.

Figure~\ref{fig:ALMA_galaxies_fgas_Mstar} presents the total gas fraction as a function of stellar mass for galaxies in A2626 (large blue markers), Virgo (turquoise markers; \citealt{chung2009, Brown2021}), and GASP (pink markers; \citealt{Moretti2023}). Within A2626, stars represent non-substructure galaxies and diamonds represent substructure galaxies. The orange dotted line indicates the median gas fraction for binned stellar masses of xCOLDGASS field galaxies \citep{Saintonge2017, Catinella2018}, serving as a reference for field environments.

Virgo galaxies generally show lower \hi and slightly reduced H$_2$ fractions, resulting in decreased total gas fractions compared to field galaxies \citep{Zabel2022}, consistent with environmental gas removal in dense clusters. In contrast, GASP jellyfish galaxies exhibit reduced \hi but enhanced H$_2$ fractions, with or without star-forming tails, maintaining total gas fractions comparable to field galaxies \citep{Moretti2023}. This suggests that ram-pressure compression may sustain their molecular gas content despite \hi depletion.

A2626 galaxies display H$_2$ fractions close to or below the xCOLDGASS field trend, while their total gas fractions remain comparable to field galaxies for both substructure and non-substructure systems. The spread around the xCOLDGASS trend suggests these galaxies are in an earlier stage of environmental processing, with \hi depletion beginning (as evidenced by disturbed \hi morphologies), but not as advanced as in Virgo. Meanwhile, H$_2$ deficiencies in A2626 are stronger than \hi deficiencies. This implies that, in contrast to the canonical picture where \hi is affected first, some A2626 galaxies may be experiencing more rapid depletion of H$_2$. Moreover, the A2626 sample includes galaxies at larger cluster-centric distances than those typically studied in Virgo, suggesting that some may be observed in the early stages of gas stripping.

The Virgo cluster, with a mass of $\sim$$5\times10^{14}$ M$_\odot$, is a dynamically unrelaxed, non-cool-core system where strong RPS leads to high \hi and moderate H$_2$ deficiencies. A2626 has a similar mass but is a cooling-core cluster \citep{Mohr1996}, indicating a more evolved and relaxed environment. SYMPHANY observations suggest that A2626 galaxies retain higher total gas fractions—driven by significantly higher \hi content—compared to Virgo, while their H$_2$ content is similar (Figure~\ref{fig:ALMA_galaxies_fgas_Mstar}). This suggests that although RPS is active in A2626, it is less intense or more gradual, allowing galaxies to preserve more of their gas reservoirs in this relatively less hostile environment.

\begin{figure}[t!]
    \centering
 {
    \includegraphics[width=0.5\textwidth]{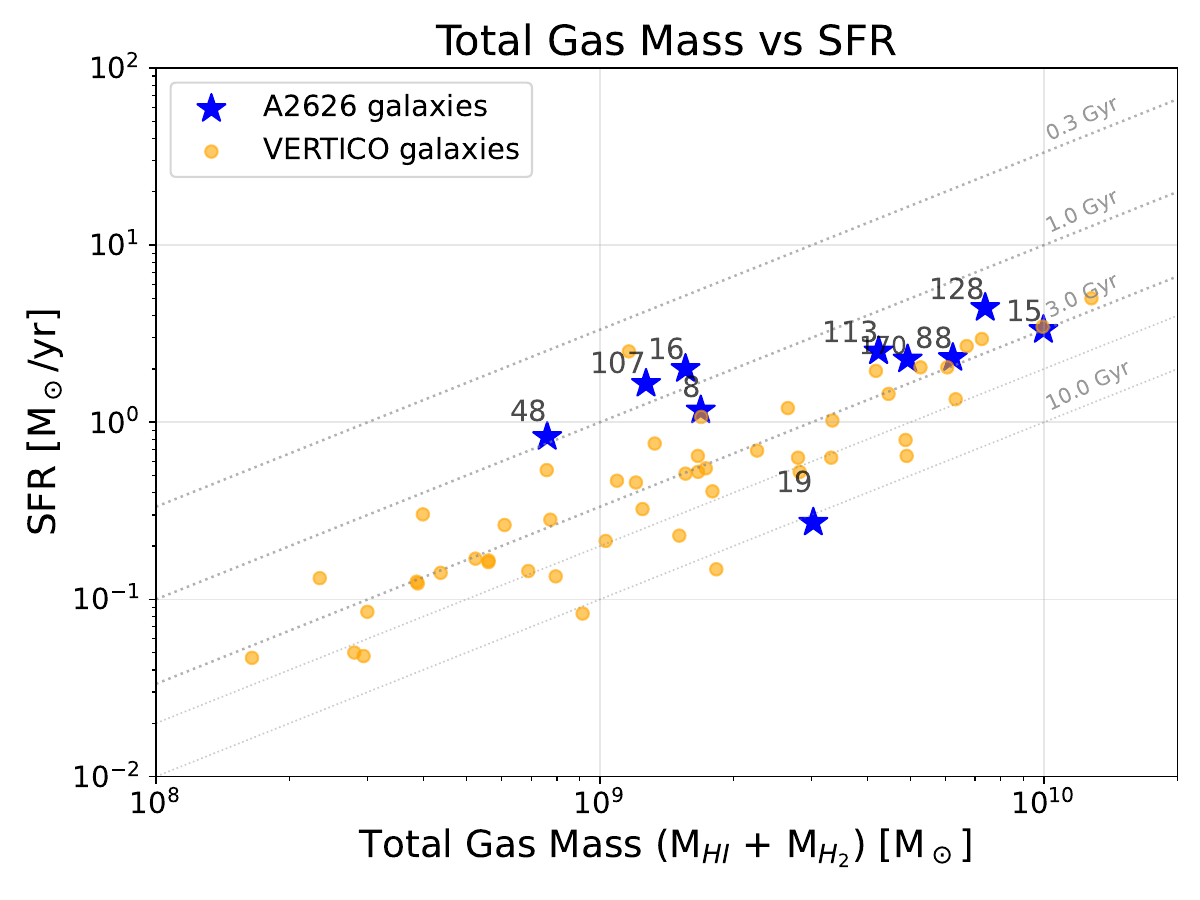} 
  } 
  \caption{Relation between total gas mass (M$\rm _{HI}$+ M$\rm _{H2}$) and SFR for galaxies in A2626 and Virgo clusters. Blue stars represent A2626 galaxies, while orange circles represent Virgo (VERTICO) galaxies. Dotted diagonal lines denote constant gas depletion times, defined as total gas mass divided by SFR. Labels indicate galaxy IDs for A2626. As expected, total gas mass and SFR are positively correlated with A2626 galaxies showing slightly elevated SFRs compared to VERTICO galaxies.} 
  \label{fig:A2626_galaxies_total_mass_SFR} 
 \end{figure}

\subsection{Total Gas Mass vs Star Formation Rate}\label{sec:total_mass_sfr}

To investigate how the cluster environment influences the evolution of galaxies, we examine the relationship between total gas mass and star formation rate (SFR) for SYMPHANY galaxies in A2626 and compare them with Virgo cluster galaxies. The Kennicutt-Schmidt (KS) scaling relation suggests a strong correlation between gas content and star formation activity, making this comparison crucial for understanding whether environmental effects alter this fundamental relation.

Figure~\ref{fig:A2626_galaxies_total_mass_SFR} presents the relationship between total gas mass (M$_{\mathrm{HI}}$ + M$_{\mathrm{H_2}}$) and SFR for galaxies in the A2626 and Virgo clusters. The total gas mass is shown on the x-axis, while the y-axis represents the SFR. Blue star markers indicate galaxies in A2626, labeled with their respective IDs, and orange circular markers represent galaxies from the Virgo (VERTICO) sample.

As expected, there is a positive correlation between total gas mass and SFR, consistent with gas fueling star formation. However, A2626 galaxies tend to exhibit higher SFRs at fixed gas mass compared to their Virgo counterparts. This offset may reflect enhanced star formation activity or a more active ISM phase in A2626 galaxies, potentially linked to differences in star formation efficiency (SFE) or environmental conditions. While both samples are biased toward star-forming systems due to selection effects (ALMA/SMA for A2626, CO detections for VERTICO), the contrast remains notable.

Compared to VERTICO galaxies, A2626 systems show moderately higher star formation efficiency. Most A2626 galaxies lie between the 0.3--3 Gyr depletion time contours, indicating that their total gas is converted into stars more rapidly than in VERTICO galaxies, which often fall below the 3 Gyr or even 10 Gyr lines. For reference, the depletion timescale of normal star-forming galaxies spans a range of $\sim$2--10 Gyr \citep{kennicutt1989, KennicuttJr.1998, bigiel2008}. The median depletion time for A2626 galaxies is $\sim$1.2 Gyr, compared to $\sim$3--5 Gyr for VERTICO, underscoring a systematically higher SFE in A2626. This suggests that, despite environmental influences, A2626 galaxies maintain efficient star formation, likely due to less severe gas stripping or more intact molecular gas reservoirs. The relatively relaxed, cool-core nature of the A2626 cluster may also contribute to preserving gas content and delaying the onset of severe depletion.

\section{Conclusions}

Our analysis of A2626 galaxies provides valuable insights into the impact of dense cluster environments on the interstellar medium (ISM). The combined \hi and CO morphological and kinematic observations confirm that atomic gas is more extended and more readily stripped, often showing truncated or one-sided profiles—a well-established signature of environmental processing. However, our results also highlight that molecular gas is not entirely shielded from these effects. Several galaxies show asymmetric CO distributions, disturbed velocity fields, and signs of H$_2$ deficiency, suggesting that molecular gas reservoirs, though more centrally concentrated, can also be disrupted or depleted in cluster environments. This points to a more complex picture of gas stripping, where both atomic and molecular phases respond to environmental forces, possibly at different stages or intensities.

Focusing on \hi and H$_2$ deficiencies, SYMPHANY galaxies generally exhibit lower deficiency values compared to Virgo galaxies, and our sample does not include any class III or IV systems typically associated with strong stripping. These confirm that environmental effects in A2626 are less extreme than in other clusters, with fewer galaxies showing significant gas depletion. The lack of correlation between H$_2$ deficiency and cluster-centric distance or velocity suggests that molecular gas evolution in A2626 may be shaped more by early infall and local interactions than by current position in the cluster. This may in part reflect the nature of A2626 as a cool-core cluster, where the denser, more stable intracluster medium (ICM) results in reduced ram pressure, potentially making galaxy-galaxy interactions more influential than RPS in shaping gas evolution.

The total gas fraction versus stellar mass analysis reveals that A2626 galaxies retain higher \hi fractions than Virgo galaxies, suggesting they are in an earlier stage of environmental processing. While Virgo’s dynamically unrelaxed state contributes to strong gas depletion via ram-pressure stripping, A2626's relaxed, cooling-core structure appears to mitigate such effects, allowing galaxies to preserve more of their gas.

A clear correlation is observed between total gas mass and SFR, underscoring the role of gas reservoirs in fueling star formation. A2626 galaxies exhibit slightly elevated SFRs and shorter depletion timescales (0.3–3 Gyr), implying moderately enhanced star formation efficiency compared to Virgo galaxies, which often fall below the 3–10 Gyr contours.

This study underscores the critical role of multi-wavelength observations in unraveling the complex environmental effects on gas content and star formation in galaxies. With additional ALMA and SMA observations now available, our expanded SYMPHANY dataset—spanning a range of stellar masses, gas content, star formation rates, and dynamical histories—will enable a more comprehensive understanding of galaxy transformation in clusters, bridging observational with theoretical models of environmental quenching.

\section*{acknowledgments}

TD acknowledges funding from an NWO Rubicon Fellowship, project number 019.231EN.001. NZ is supported through the South African Research Chairs Initiative of the Department of Science and Technology and National Research Foundation. My acknowledgements: YLJ acknowledges support from the Agencia Nacional de Investigaci\'on y Desarrollo (ANID) through Basal project FB210003, FONDECYT Regular projects 1241426 and 123044, and  Millennium 
Science Initiative Program NCN2024\_112. This paper makes use of the following ALMA data: ADS/JAO.ALMA\#2021.1.01375.S ALMA is a partnership of ESO (representing its member states), NSF (USA) and NINS (Japan), together with NRC (Canada), MOST and ASIAA (Taiwan), and KASI (Republic of Korea), in cooperation with the Republic of Chile. The Joint ALMA Observatory is operated by ESO, AUI/NRAO and NAOJ. The National Radio Astronomy Observatory is a facility of the National Science Foundation operated under cooperative agreement by Associated Universities, Inc. This paper makes use of the observations from the following SMA project: 2021A-SO65. The Submillimeter Array is a joint project between the Smithsonian Astrophysical Observatory and the Academia Sinica Institute of Astronomy and Astrophysics and is funded by the Smithsonian Institution and the Academia Sinica.

\vspace{5mm}
\facilities{ALMA, SMA, MeerKAT}

\software{astropy \citep{2013A&A...558A..33A,2018AJ....156..123A}
          }

\bibliography{refs}{}
\bibliographystyle{aasjournal}

\end{document}